\documentclass{ws-procs9x6}

\begin{document}

\title{Exclusive diffractive final states in electron-proton collisions
%Instructions for Producing a Camera-Ready Manuscript 
%using Latex for Publication in Conference 
%Proceedings\footnote{\uppercase{T}his work is supported by etc, etc.}
}

\author{Salvatore Fazio
%C.~E. JIM and O. SPINAS\footnote{\uppercase{W}ork partially
%supported by grant 2-4570.5 of the \uppercase{S}wiss 
%\uppercase{N}ational \uppercase{S}cience \uppercase{F}oundation.}
}

\address{Calabria University and INFN Cosenza,\\
Department of Physics, ponte P. Bucci- cubo 31C,\\
I-87036, Arcavacata di Rende (Cosenza), Italy 
%World Scientific Publishing Co., Inc, \\
%1060 Main Street, \\ 
%River Edge, NJ 07661, USA\\ 
%E-mail: wspc@wspc.com
}

%\author{T.~R. SIMON, S. CLARKE and S.~N. GERALD}

%\address{World Scientific Publishing Co Ltd, \\ 
%57 Shelton Street, \\
%London WC2H 9HE, England\\
%E-mail: wspc@wspc.ox.uk}  

\maketitle

\abstracts{
The exclusive diffractive production of vector mesons and real photons in $ep$ collisions has been studied at HERA in a wide kinematic range. The most recent experimental results are presented. 
%%with a particular emphasis on the measurements of the diffractive cross section as a function of the transferred four-momentum at the proton vertex, $t$. 
%This is where the abstract should be placed. It should consist of one paragraph giving a concise summary of the material in the article below.  Replace the title, authors, and addresses within the curly brackets with your own title, authors, and addresses. You may have as many authors and addresses as you like. It is preferable not to use footnotes in the abstract or the title; the acknowledgments for funding bodies etc. are to be placed in a separate section at the end of the text. Please see the appendices too.
}

\section{Introduction}
The diffractive scattering is a process where the colliding particles scatter at very small angles and without any color flux in the final state. This involves a propagator carrying the vacuum quantum numbers, called Pomeron and described, in the soft regime, within the Regge theory. The discovery of a big amount of diffractive events in DIS regime provided a hard scale which can be
varied over a wide range and therefore it is an ideal testing for QCD models.

In particular, the diffractive production of Vector Mesons (VMs) and real photons in $ep$ collisions allows to study the transition from the soft to the hard regime in strong interactions. The hard regime (high energy and low Bjorken-$x$) is dominated by the exchange of a hard Pomeron well described by perturbative QCD (pQCD), while at low-$x$ the interaction is well described within the Regge phenomenology. Indicating with $Q^2$ the virtuality of the exchanged photon and with $M^2$ the square mass of the produced VM, HERA data suggested a universal hard scale, $Q^2+M^2$, for the diffractive exclusive pruduction of VM and real photons, which indicates the transition from the soft to the hard regime.
%Since the first operation period in 1992, ZEUS and H1, the two experiments dedicated to the DIS physics at HERA, observed that an amount $(\sim 10 \%)$ of lepton-proton DIS events had a diffractive origin opening a new area of studies in diffractive production mechanism. 
%providing a hard scale which can be varied over a wide range and therefore it is an ideal testing for QCD models of diffractive scattering.

%The diffractive production of vector mesons (VM) and real photons in $ep$ interactions leads to the extraction of the Generalized Parton Distribution functions (GPDs), containing combined imformations about the longitudinal momentum distribution of partons and their position on the trasfers plain. The GPD-based calculations will be very helpfull in the description of the Higgs boson diffractive production mechanism, which will be experimentally studied with the LHC accelerator. 

\section{$Q^2$ and $W$ dependence of the cross section}

\begin{figure}[htbp]
%\epsfxsize=10cm   %width of figure - will enlarge/reduce the figures
%\epsfbox{fig3.eps}
%\figurebox{2cm}{3cm}{} %to have a box alone 
\centering
\includegraphics[width=0.7\textwidth,angle=0]{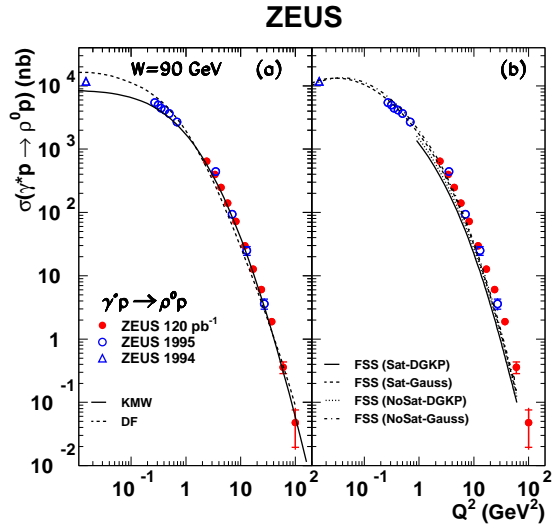}
%%\centerline{\epsfxsize=4.1in\epsfbox{W_php.eps}}   
\caption{The $\gamma^*p\rightarrow\rho^0p$ cross section as a function of $Q^2$ measured at $W=90\;GeV^2$ and comared in (a) and (b) with different models as described in the text.  \label{q2_rho}}
\end{figure}

A new precision measurement of the reaction $\gamma^*p\rightarrow\rho^0 p$ was published by ZEUS~\cite{zeus_rho}. It was found that the cross section falls steeply with the increasing of $Q^2$ but, unlike it was observed for the $J/\psi$ electroproduction~\cite{zeus_jpsi,h1_jpsi}, it cannot be described by a simple propagator term like $\sigma\propto (Q^2+M^2)^{-n}$, in particular an $n$ value increasing with $Q^2$ appears to be favored. Figure~\ref{q2_rho} reports the cross section for the $\rho^0$ electroproduction versus $Q^2$ compared with several theoretical predictions: the KWM model~\cite{KMW} based on the saturation model, the FSS model~\cite{FSS} with and without saturation and the DF model~\cite{DF}. None of the available models gives a good description of the data over the full kinematic range of the measurement.

%\section{$W$ dependence}

\begin{figure}[htbp]
%\epsfxsize=10cm   %width of figure - will enlarge/reduce the figures
%\epsfbox{fig3.eps}
%\figurebox{2cm}{3cm}{} %to have a box alone 
\centering
\includegraphics[width=0.6\textwidth,angle=0]{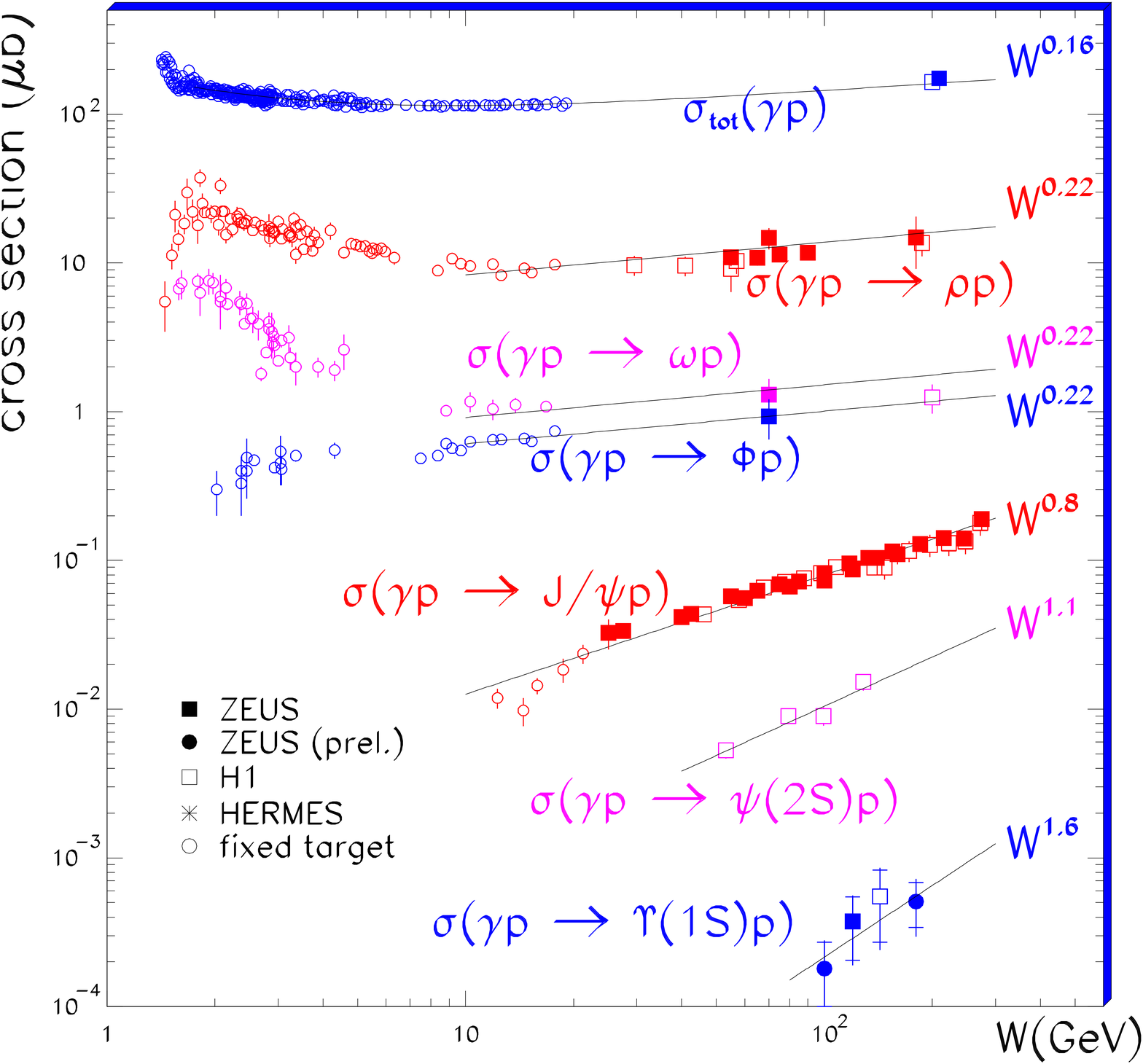}
%%\centerline{\epsfxsize=4.1in\epsfbox{W_php.eps}}   
\caption{The $W$ dependence of the cross section for exclusive VM photoproduction together with the total photoproduction cross section. Lines are the result of a $W^{\delta}$ fit to the data at high $W$-energy values. \label{W_php}}
\end{figure}

The soft to hard transition can be observed looking at the dependence of the VM photoproduction ($Q^2=0$) cross section from the $\gamma^*p$ centre of mass energy, $W$, where the scale is provided by $M^2$. Figure~\ref{W_php} collects the $\sigma ( \gamma^* p\rightarrow Vp )$ as a function of $W$ from the lightest vector meson, $\rho^0$, to the heaviest, $\Upsilon$, compared to the total cross section. 
The cross section rises with the energy as $W^{\delta}$, where the $\delta$ exponent increases with the hard scale $M^2$ as expected for a transition from the soft to the hard regime. New results on the $\Upsilon$ photoproduction~\cite{upsilon}, recently published by ZEUS, confirmed the steeper rise of $\sigma(W)$ for higher vector meson masses. 

\begin{figure}[htbp]
%\epsfxsize=10cm   %width of figure - will enlarge/reduce the figures
%\epsfbox{fig3.eps}
%\figurebox{2cm}{3cm}{} %to have a box alone 
\centering
\begin{tabular}{cc}
\includegraphics[width=0.5\textwidth,angle=0]{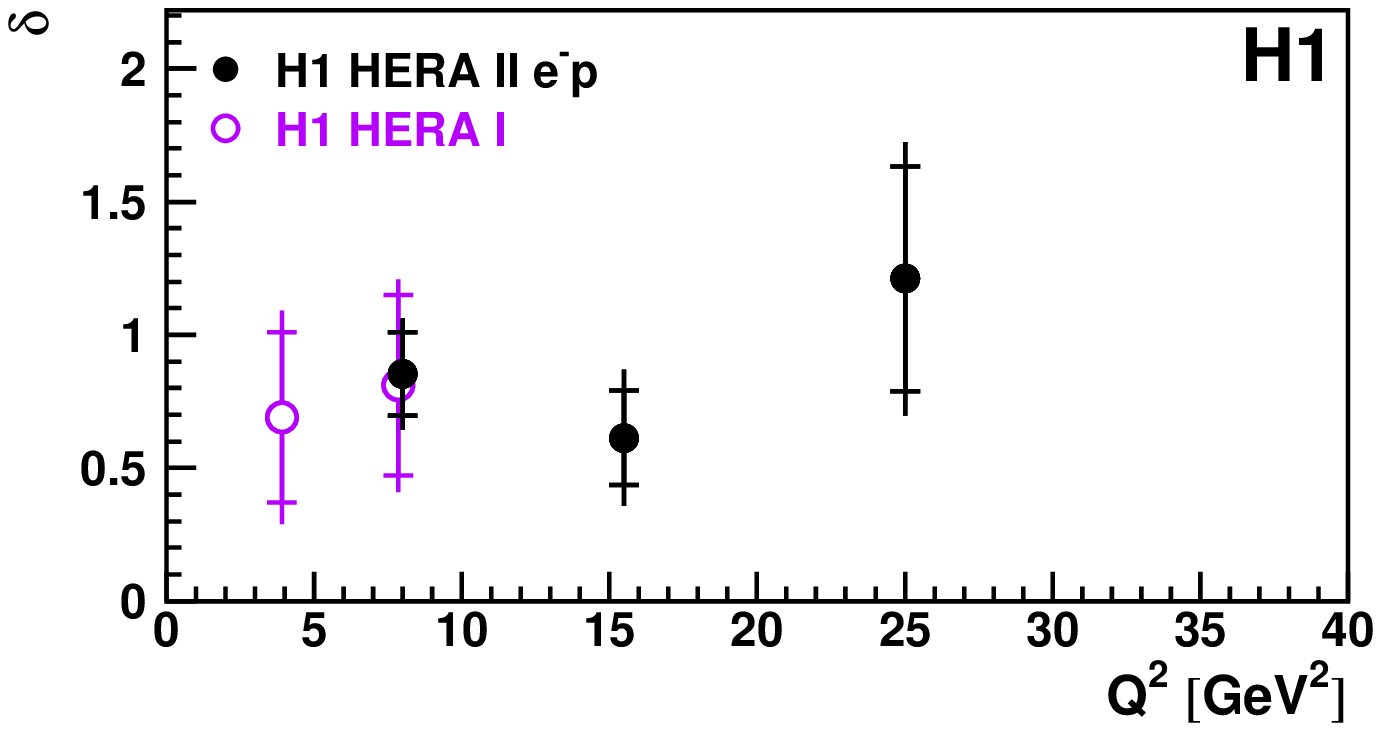}
\includegraphics[width=0.5\textwidth,angle=0]{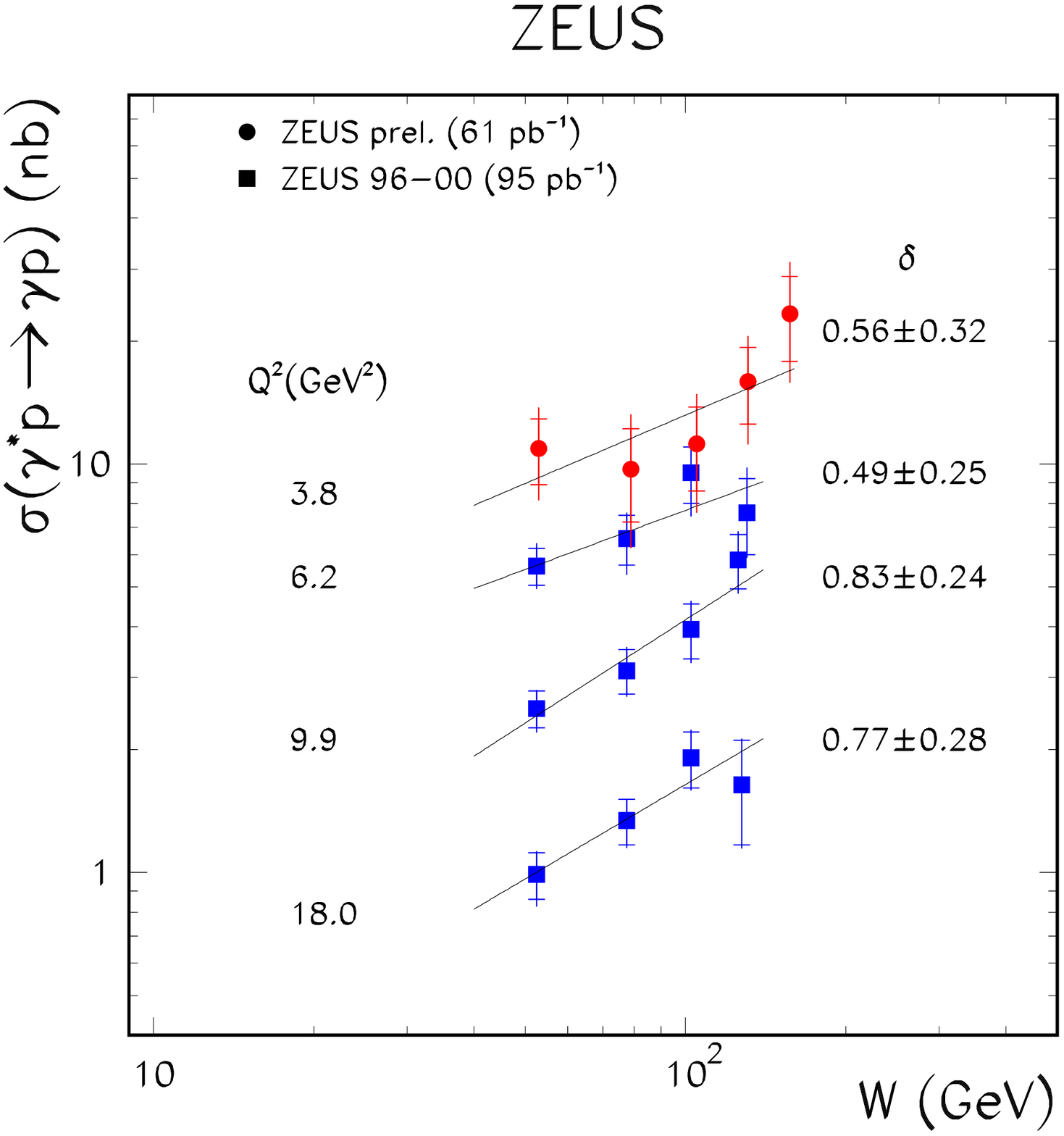}
\end{tabular}  
%%%%\includegraphics[width=0.8\textwidth,angle=0]{W_dvcs_collec.eps}
%%\centerline{\epsfxsize=4.1in\epsfbox{W_dvcs_collec.eps}}   
\caption{The $W$ dependence of the cross section for a DVCS process. Lines come from a $W^{\delta}$ fit to the data. Left: the H1 measurement of the $\delta$ slope as a function of $Q^2$. Right: the new ZEUS preliminary measurement at low $Q^2$ (dots) together with the published measurements (squares). \label{W_dvcs}}
\end{figure}

The transition from the soft to the hard regime can also be studied varying $Q^2$. Recent results were achieved by H1~\cite{h1_dvcs} and ZEUS~\cite{zeus_dvcs} for the exclusive production of a real photon, the Deeply Virtual Compton Scattering (DVCS), where the hard scale is provided only by the photon virtuality, $Q^2$. Figure~\ref{W_dvcs} shows the H1 (left) and the ZEUS (right) results. The steep rise with $W$ of the cross section even at low-$Q^2$, seems to suggest that the most sensitive part to the soft scale comes from the wave function of the pruduced VM. A similar result was obtained for the $J/\psi$ electroproduction~\cite{zeus_jpsi,h1_jpsi}.
 
\begin{figure}[htbp]
%\epsfxsize=10cm   %width of figure - will enlarge/reduce the figures
%\epsfbox{fig3.eps}
%\figurebox{2cm}{3cm}{} %to have a box alone 
\centering
\begin{tabular}{cc}
\includegraphics[width=0.53\textwidth,angle=0]{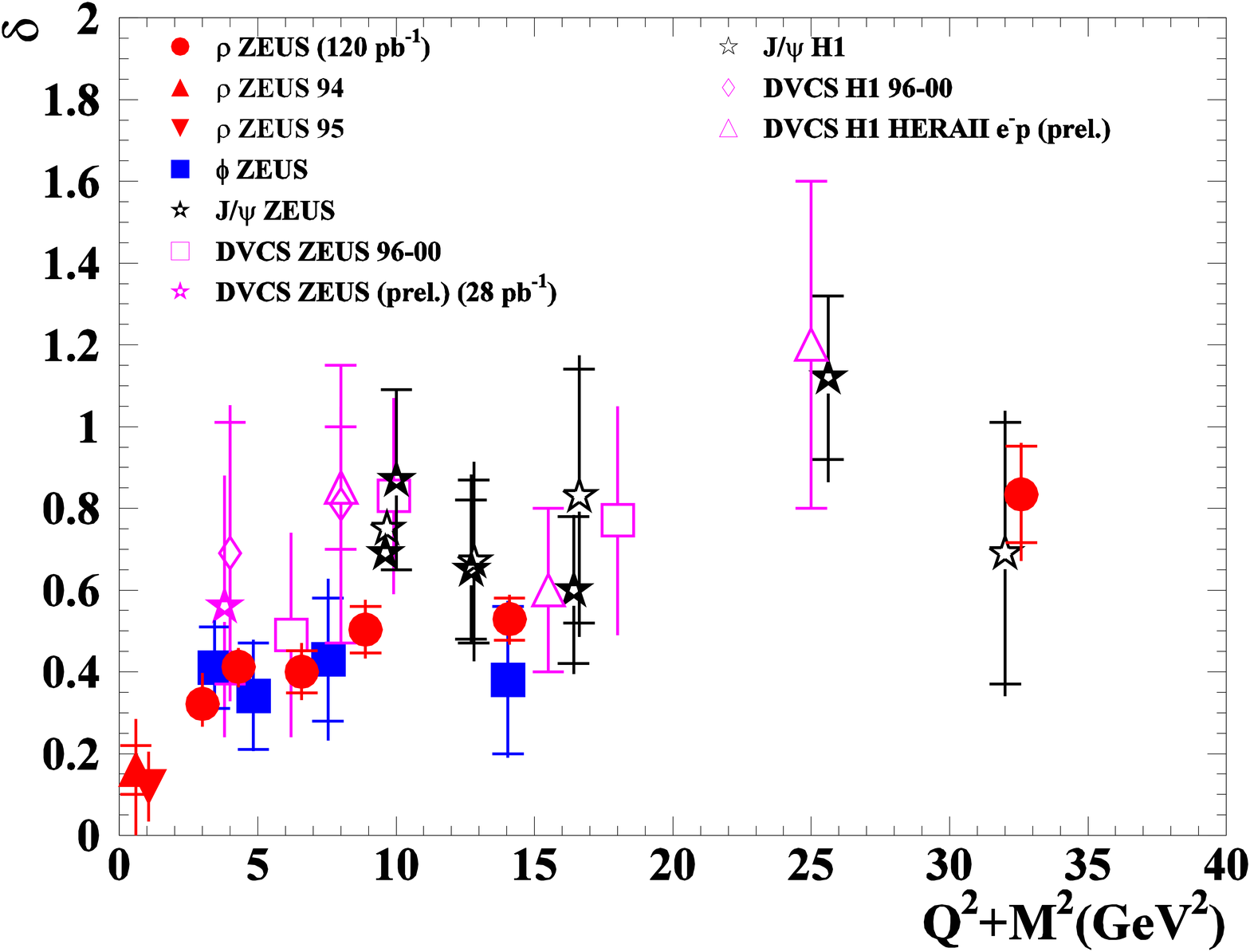}
%%\centerline{\epsfxsize=4.1in\epsfbox{W_php.eps}}
\includegraphics[width=0.46\textwidth,angle=0]{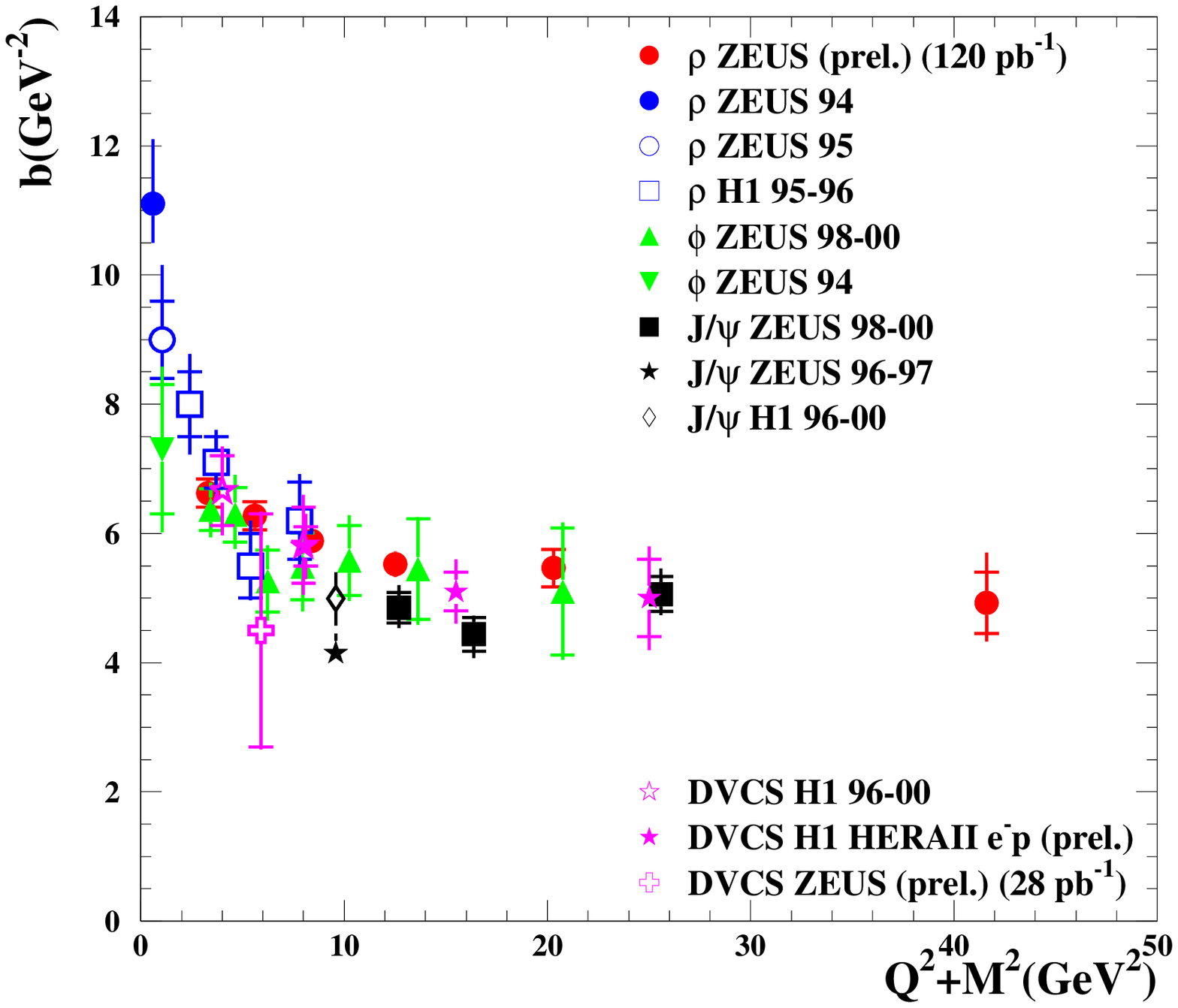}
\end{tabular}   
\caption{The dependence on the hard scale $Q^2+M^2$ of the value $\delta$ (left) extracted from a fit $W^{\delta}$ and of the slope $b$ (right) extracted from a fit $\frac{d\sigma}{dt}\propto e^{b|t|}$ for the exclusive VM electroproduction. DVCS is also included. \label{W_dis}}
\end{figure}

The electroproduction of a large variety of VMs was studied at different $Q^2$ values and the corresponding slope $\delta$ is reported in Fig.~\ref{W_dis} (left) versus the scale $Q^2+M^2$, including the DVCS measurements. Data show a logarithmic shape $\sigma\propto \ln(Q^2+M^2)$ and the behaviour seems to be universal with $\delta$ increasing from 0.2 at low scale, as expected from a soft Pomeron exchange~\cite{Wsoft} to $\sim 0.8$ at large scale values. 

\section{$t$ dependence of the cross section}

The differential cross section as a function of $t$ can be parametrised by a fit $\frac{d\sigma}{d|t|}\propto e^{b|t|}$. Figure~\ref{W_dis} (right) reports the collection of the $b$ values versus the scale $Q^2+M^2$ for the electroproduction of VMs and DVCS, with $b$ decreasing from $\sim 11\; GeV^{-2}$ to $\sim 5\; GeV^{-2}$ as expected in hard regime.  

%%%\begin{figure}[htbp]
%\epsfxsize=10cm   %width of figure - will enlarge/reduce the figures
%\epsfbox{W_php.eps}
%\figurebox{2cm}{3cm}{W_php.eps} %to have a box alone 
%%%\centering
%%%\includegraphics[width=0.5\textwidth,angle=0]{beps07.eps}
%\centerline{\epsfxsize=4.1in\epsfbox{W_php.eps}}   
%%%\caption{The dependence on the hard scale $Q^2+M^2$ of the slope $b$ extracted from a fit $\frac{d\sigma}{dt}\propto e^{b|t|}$ for the exclusive vector-meson electroproduction. DVCS is also included. \label{b_dis}}
%%%\end{figure}

\begin{figure}[htbp]
\centering
\begin{tabular}{cc}
\includegraphics[width=0.5\textwidth,angle=0]{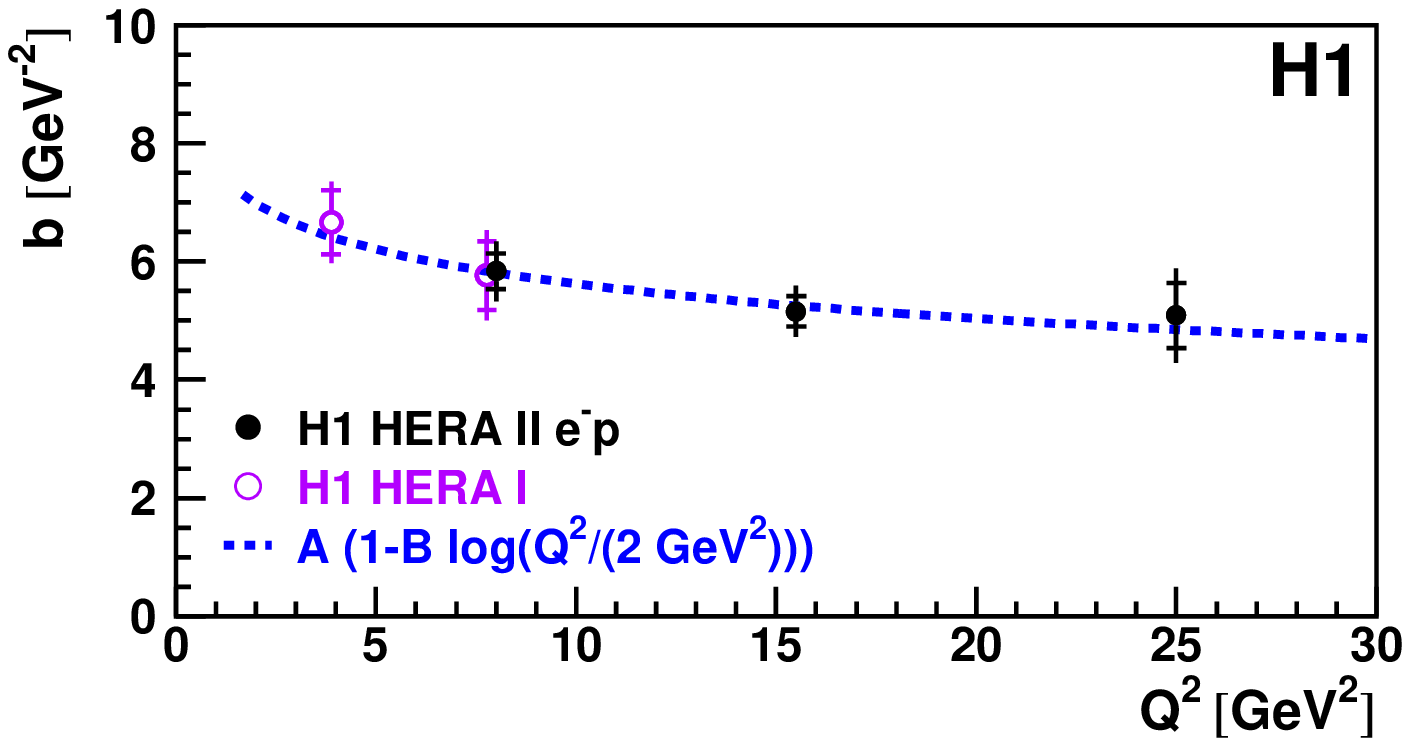}
\includegraphics[width=0.5\textwidth,angle=0]{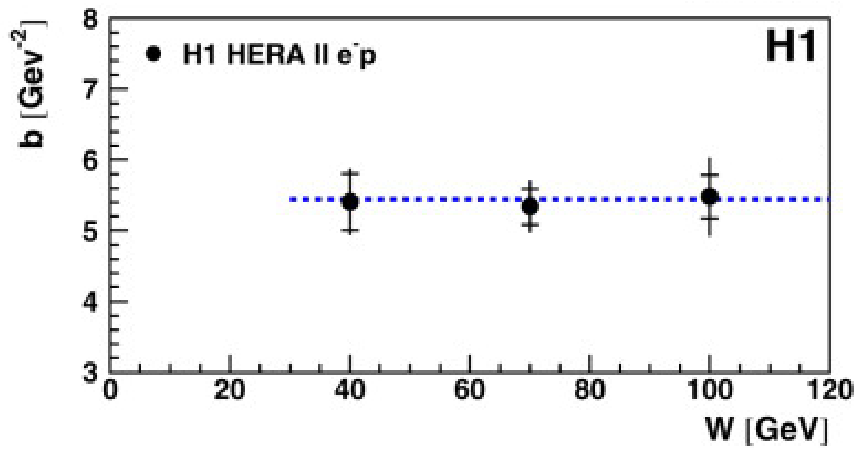}
\end{tabular}
\caption{The $t$ slope parameter $b$ as a function of $Q^2$ (left) and $W$ (right).}
\label{b_dvcs}
\end{figure}

The measurement of $d\sigma/d|t|$ for the DVCS process, recentrly published by the H1 Collab~\cite{h1_dvcs}, where $t$ was obtained from the transverse momentum distribution of the photon, studied $b$ versus $Q^2$ and $W$ as shown in Fig.~\ref{b_dvcs}. $b$ seems to decrease with $Q^2$ up to the value expected for a hard process but it doesn't depend on $W$.   
A new preliminary ZEUS measurement~\cite{zeus_dvcs} of $d\sigma/d|t|$ has been achieved from a direct measurement of the proton final state of using a spectrometer based on the roman pot thechnique. The result $b=4.4\pm 1.3~(stat.)\pm 0.4~(syst.)~GeV^{-2}$, measured at $Q^2=5.2~GeV^2$ and $W=104~GeV$, is consistent, within the large uncertainties due to the low acceptance of the spectrometer, with the H1 result~\cite{h1_dvcs} of $b=5.45\pm 0.19~(stat.)\pm 0.34~(syst.)~GeV^{-2}$ at $Q^2=8~GeV^2$ and $W=82~GeV$.

Since $b$ value can be related via a Fourier transform to the impact parameter and assuming that the exclusive process in the hard regime is dominated by gluons, the relation $\langle r^2\rangle=b(\hbar c)^2$ can be used to obtain the radius of the gluon confinement area in the proton. $b\sim 5\;GeV^2$ corresponds to $\langle r^2\rangle\sim 0.6\;fm$ smaller than the proton radius ($\sim 0.8\;fm$) indicating that the gluons are well contained within the charge-radius of the proton.

\end{document}